\documentclass[11pt]{article}
\usepackage{cs12,html}
\usepackage{graphicx,hyperref}

\begin{document}
 
\title{A Study of the Coronal Plasma in RS CVn binary systems}
 
\author{M. Audard\altaffilmark{1}, %
M. G\"udel\altaffilmark{1}, %
A. Sres\altaffilmark{1,2}, %
R. Mewe\altaffilmark{3}, %
A.~J.~J. Raassen\altaffilmark{3,4}, %
E. Behar\altaffilmark{5}, %
C.~R. Foley\altaffilmark{6}, %
R.~L.~J.~van~der~Meer\altaffilmark{3}
}
\altaffiltext{1}{Paul Scherrer Institut, CH}
\altaffiltext{2}{Institute of Astronomy, ETHZ, CH}
\altaffiltext{3}{Space Research Organization of the Netherlands, NL}
\altaffiltext{4}{Astronomical Institute ``Anton Pannekoek'', NL}
\altaffiltext{5}{Columbia Astrophysics Laboratory, USA}
\altaffiltext{6}{Mullard Space Science Laboratory, UK}

\index{*HR 1099}
\index{*UX Ari}
\index{*$\lambda$ And}
\index{*Capella}

\index{X-ray}
\index{XMM-Newton}
\index{RGS}
\index{Coronae!abundances}
\index{Abundances}
\index{Activity}

\begin{abstract}
\textit{XMM-Newton} has been performing comprehensive 
studies of X-ray bright RS CVn binaries in its Calibration and
Guaranteed Time programs. We present results from ongoing investigations 
in the context of a systematic study of coronal emission from RS CVns. 
We concentrate in this paper on coronal abundances and
investigate the abundance pattern in RS CVn binaries as a function
of activity and average temperature. A transition from an Inverse
First Ionization Potential (FIP) effect towards an absence of a clear trend
is found in intermediately active RS CVn systems. This scheme corresponds well into the long-term
evolution from an IFIP to a FIP effect found in solar analogs. We further study 
variations in the elemental abundances during a large flare. 
\end{abstract}

\section{Introduction}
The RS CVn binary system class is loosely defined as consisting of a chromospherically
active evolved star orbiting within a few days around a main-sequence or subgiant companion
(Hall 1976). Their short rotation period implies high levels of activity (e.g., 
Noyes et al. 1984), observed as strong emission of 
chromospheric lines and saturated X-ray emission (Dempsey et al. 1993).
The high-resolution X-ray spectra of the brightest and nearby RS CVn binary systems 
obtained by  \textit{XMM-Newton} are well-exposed and
provide a high signal-to-noise ratio. Abundant coronal elements (C, N, O, Ne,
Mg, Si, S, Ar, Ca, Fe, and Ni) produce a rich spectrum of electronic transitions
in the EUV and X-rays, allowing us to perform benchmark studies of atomic databases. 
Recent results with \textit{Chandra} and \textit{XMM-Newton} show that models reproduce
the observed spectra fairly well (e.g., Audard et al. 2001a, Behar, Cottam, 
\& Kahn 2001), although a significant number of lines, mainly from Si, S, Ar, and Ca
L-shell lines, are either absent in the atomic databases or are not properly 
reproduced (Audard et al. 2001a). In this study, we have therefore discarded
wavelength ranges where such lines dominate in order not to bias the convergence
of the spectral fits, and especially to get more accurate elemental abundances.

Past stellar coronal abundance determinations have been done with CCD spectra of
moderate spectral resolution (e.g., Drake 1996, G\"udel et al. 1999) or with low sensitivity 
spectrometers (e.g., Drake, Laming, \& Widing 1995, Laming, Drake, \& Widing 
1996, Schmitt et al. 1996, Drake, Laming,  \& Widing 1997). Abundance studies 
of stellar coronae are a powerful means to better understand the 
well-studied, but still puzzling, abundance pattern in the Sun: in brief, the 
solar corona and the solar wind display a so-called ``First Ionization Potential'' (FIP) 
effect, for which the current consensus is that the abundances of low-FIP ($<10$~eV) 
elements are \emph{enhanced} relative to their respective photospheric abundance, 
while the abundances of high-FIP ($>10$~eV) elements are photospheric 
(e.g., Haisch, Saba, \& Meyer 1996). Stellar coronal spectra often showed a 
metal abundance deficiency relative to the \emph{solar}
photospheric abundances (Schmitt et al. 1996), with Fe/Fe$_\odot \approx 0.1 -
0.2$ in active RS CVn binary systems. More detailed studies with 
\textit{EUVE} showed either the absence of any FIP-related bias (Drake et al. 1995), 
or the presence of a FIP effect  in inactive stellar coronae (Drake et al. 1997). 
The current generation of X-ray observatories,  \textit{XMM-Newton} and \textit{Chandra}, 
combine high spectral resolution with moderate effective areas to routinely 
obtain excellent data to measure abundances in stellar coronae.

An analysis of a deep exposure of the \textit{XMM-Newton} RGS spectrum (Brink\-man et al. 2001) 
of the RS CVn binary system HR~1099 showed a trend towards enhanced
high-FIP elemental abundances (normalized to O and relative to the solar
photospheric abundances, Anders \& Grevesse 1989), while low-FIP elemental
abundances were depleted; this effect was dubbed the ``Inverse FIP'' effect.
Different active stars also show such a trend (G\"udel et al. 2001ab), 
while the intermediately active binary Capella displays neither a FIP nor
an IFIP effect (Audard et al. 2001a). It is practice to
normalize stellar coronal abundances to the \emph{solar} photospheric abundances, 
while they should better be normalized to the \emph{stellar} photospheric 
abundances. The latter are, however, hard to measure due to enhanced chromospheric activity, 
high rotation rate and the presence of spots in active stars, particularly in RS CVn
binaries. Nevertheless, for some stars, photospheric abundances are known. 
G\"udel et al. (2002; also in these proceedings) discuss the transition from an
inverse FIP effect to a ``normal'' FIP effect in the long-term coronal evolution 
from active to inactive solar analogs; all targets have photospheric abundances 
indistinguishable from those of the Sun, which suggests that the observed 
transition is real.

While low-FIP elements appear underabundant in the quiescent state, a notable
increase of the abundances of ``metals'' during flares is often observed
(e.g., Ottmann \& Schmitt 1996, Favata et al. 2000). More precisely,
time-dependent spectroscopy of a large flare in UX Ari (G\"udel et al. 1999)
showed that low-FIP elements increased more significantly than the high-FIP 
elements. A recent high-resolution X-ray spectroscopic study of
a flare in HR~1099 with \textit{XMM-Newton} showed a similar behavior
(Audard, G\"udel, \& Mewe 2001b).

In this ``electronic'' paper, we present a study of abundances in RS CVn binary 
systems observed by  \textit{XMM-Newton}. In brief, it shows i) a 
transition from an inverse FIP effect to an absence of a FIP bias with 
decreasing activity, compatible with a similar transition observed in solar 
analogs (G\"udel et al. 2002; also in these proceedings), ii) a depletion of 
low-FIP elemental coronal abundances with increasing average coronal 
temperature, while high-FIP elemental abundances stay constant, iii) 
an enhancement of low-FIP elemental abundances during flares, while
high-FIP elemental abundances again stay constant.

\section{Observations and Data Analysis}
\begin{figure}[!t]
\includegraphics[width=\textwidth]{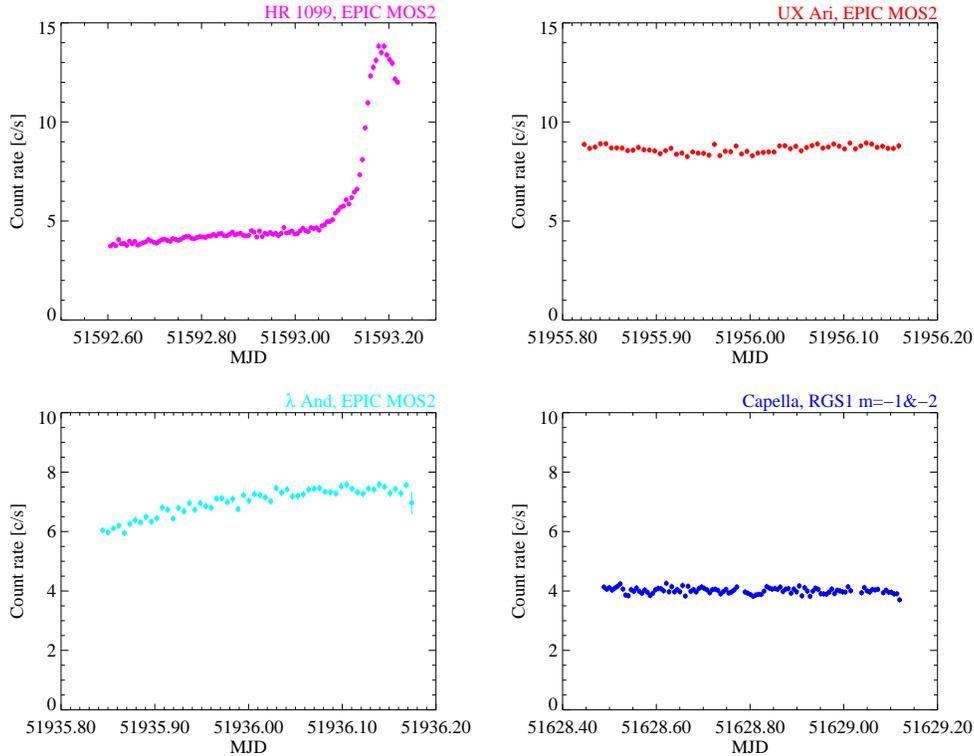}
\caption{X-ray light curves of HR~1099, UX Ari, $\lambda$ And, and Capella
binned at 500~s. 
EPIC MOS2 light curves are shown, except for Capella where the sum of the RGS1 
first and second order light curves is given, due to substantial pile-up and
optical contamination of the EPIC data. Note that the EPIC count rate for
HR~1099 is smaller as we had to use an
annulus extraction region to account for pile-up in the
central part of the PSF. Also, note that only the quiescent
part of HR~1099 has been included in the analysis; see Fig.~6 for 
the flare analysis (also Audard et al. 2001b).}
\end{figure}

\begin{figure}
\includegraphics[width=\textwidth]{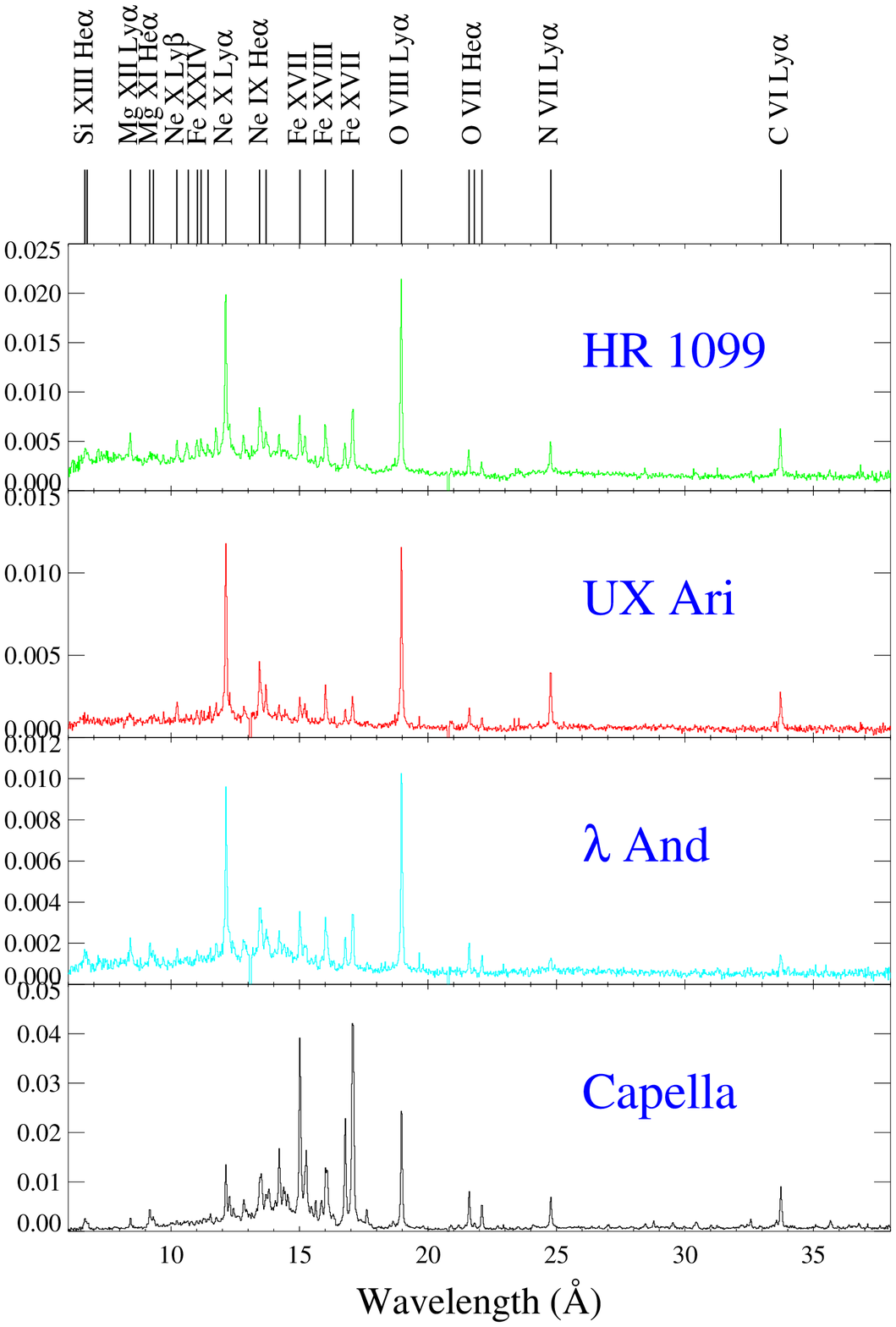}
\caption{Fluxed \textit{XMM-Newton} RGS spectra with some line identifications.
The spectra are binned to a resolution of 25 m\AA.}
\end{figure}

\textit{XMM-Newton} observed a number of RS CVn binary systems as part of the 
RGS stellar Guaranteed Time Program (see G\"udel et al. in these proceedings). 
Here, we present an analysis of the quiescent observations of
HR~1099, UX Ari, $\lambda$~And, and Capella. Their light curves are shown in
Figure~1 and their high-resolution RGS spectra in Figure~2. The RGS1, RGS2, and 
EPIC MOS2 spectra (except for Capella where no EPIC data are available) were 
fitted simultaneously in XSPEC 11.0.1aj (Arnaud 1996) with the
\texttt{vapec} model. We have removed significant
parts of the RGS spectra above 20~\AA\  to take into account the incompleteness and
inaccuracy of the atomic database for non-Fe L-shell transitions.
Additionally, some Fe L-shell lines with inaccurate atomic data were not fitted. 
A free multiplicative constant model has been introduced for cross-calibration uncertainties, finite
extraction region and, in the case of HR~1099, an annulus-shaped extraction region.
Notice that RGS1 and RGS2 each suffer from the loss of one CCD, but the
combined spectra cover the whole RGS wavelength range (den Herder et al.
2001). Here we present results from fits on a grid of 10 components with 
fixed T, but free EM and abundances (the latter linked between the components).

\section{Results}

\begin{figure}[!t]
\includegraphics[width=\textwidth]{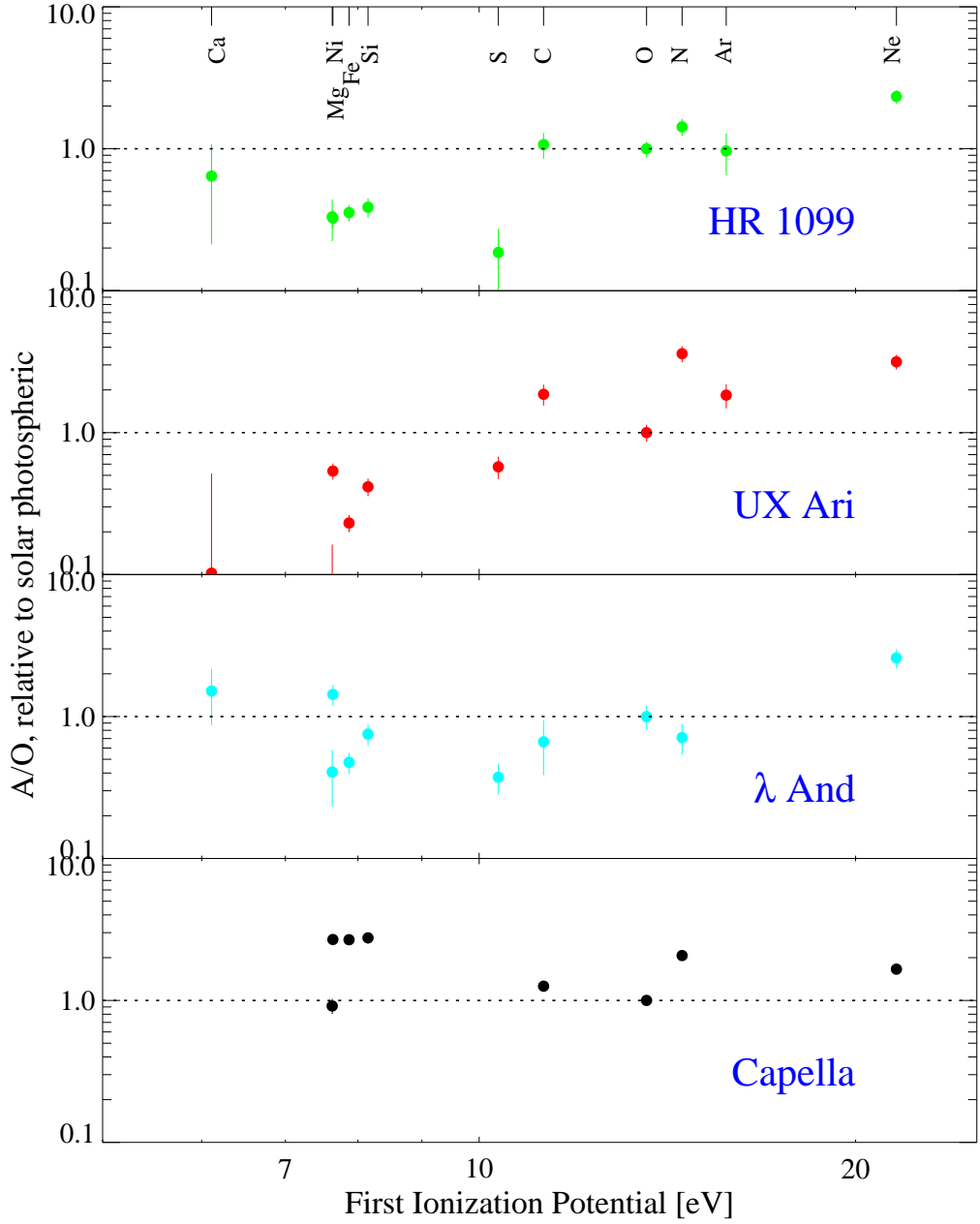}
\caption{Coronal abundance ratios in RS CVn binaries as a function of the First
Ionization Potential. The abundances have been normalized to O to allow 
for comparison. We used solar photospheric abundances from 
Anders \& Grevesse (1989), except for Fe (Grevesse \& Sauval 1999).}
\end{figure}

Figure~3 shows the coronal abundances normalized to the O abundance in order to ease
comparison between the different stars. We used the solar photospheric abundances from Anders \&
Grevesse (1989), except for the Fe abundance which was taken from Grevesse \&
Sauval (1999). The panels are ordered in decreasing activity from top to
bottom. Clearly, ratios for high-FIP elements exceed the ratios for low-FIP
elements in very active stars (HR~1099, UX Ari), while for the less
active, not tidally locked $\lambda$ And, a poor correlation is observed.
In the intermediately active Capella, no FIP bias is present although a
possible weak FIP effect could be suggested. However, any bias is definitive
only if
coronal abundances of stars are compared to their respective photospheric
abundances rather than to the solar values. Unfortunately, few photospheric abundances are 
known for RS CVn binaries. $\lambda$ And is an exception: we have used 
the abundances derived by Donati, Henry, \& Hall (1995) to normalize our 
coronal abundances (Fig.~4). Again, no clear correlation can be observed. 
Such a procedure needs to be applied to stars that do show a clear IFIP bias
when normalized to solar photospheric values. New accurate measurements of
photospheric abundances in these objects are timely.

The ``average'' temperature of a stellar corona is an indicator of activity that
is defined here as $\log <T> = (\Sigma_i \log T_i \times EM_i) /(\Sigma_i EM_i)  $,
where $T_i$ and $EM_i$ are
the 10-T model temperatures and emission measures, respectively. We present
samples of the abundance ratios relative to O (for the low-FIP Fe and the 
high-FIP Ne) as a function of $<T>$ (Fig.~5). 
Data points from solar analogs (G\"udel et al. 2002; in these proceedings) 
have been added; the panels show a very different behavior: while the Fe/O 
ratios exponentially decrease with increasing temperature, the Ne/O ratios 
show no correlation with the average coronal temperature. Similar behavior is
observed in other low-FIP and high-FIP elements, respectively.

\begin{figure}
\includegraphics[width=\textwidth]{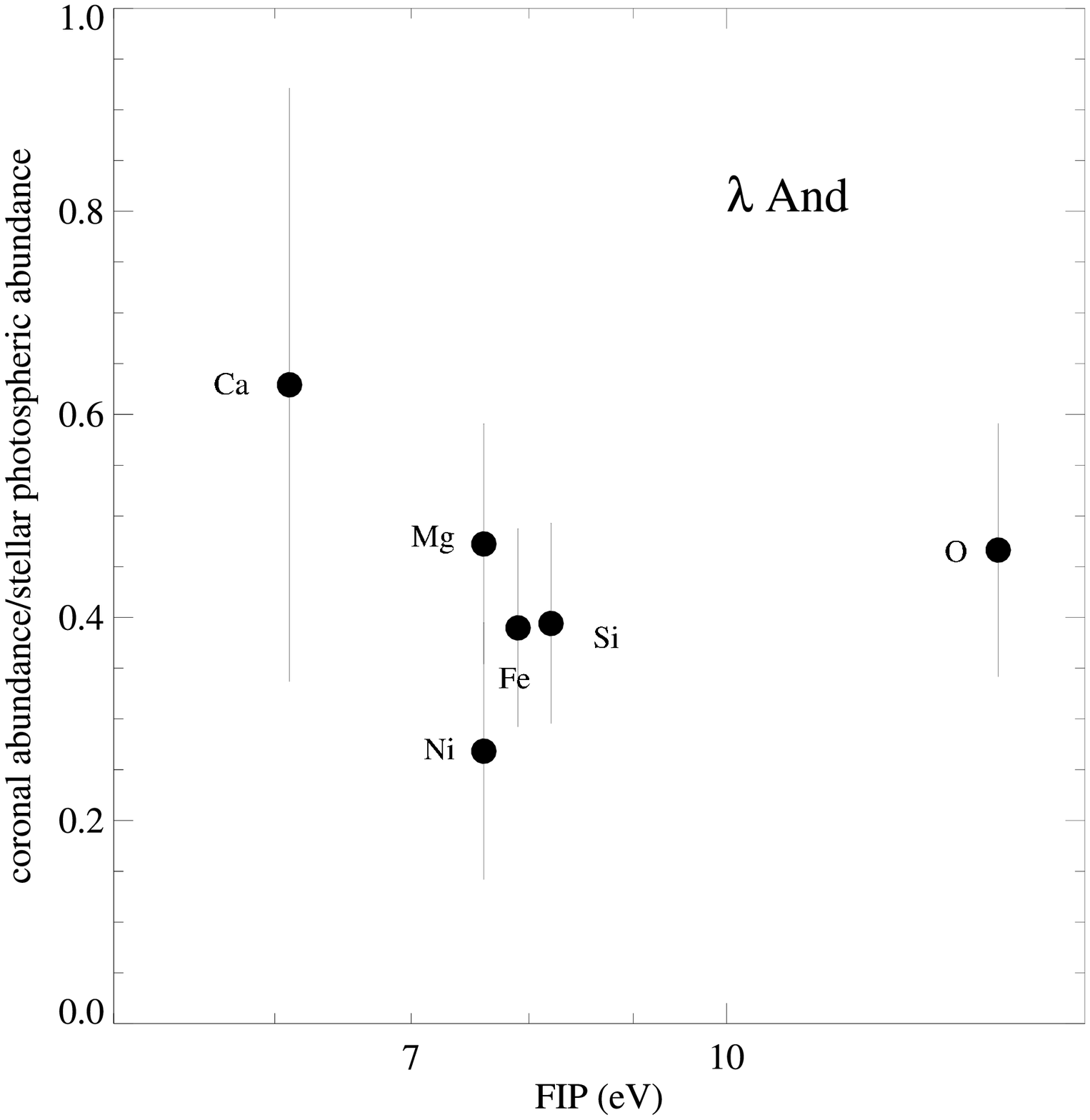}
\caption{Coronal abundances of the RS CVn binary $\lambda$ And relative to 
\emph{stellar} photospheric abundances (Donati et al. 1995) as a function of FIP.}
\end{figure}

While the preceding results are valid fo a quiescent state, previous data showed
that the average metallicity Z or the Fe abundance can increase during large
flares. Medium spectral resolution observations with \textit{ASCA} allowed G\"udel et al.
(1999) to obtain time-dependent measurements of several elemental abundances
during a large flare in UX Ari. They found that low-FIP elements 
increased more significantly than the high-FIP elements, although the latter
could not be well-constrained due to blending (Ne) or low signal (Ar). A similar
behavior was observed with recent \textit{XMM-Newton} data of a flare in HR 1099
(Audard et al. 2001b). Here we present results of a reanalysis of this flare
applying a more recent calibration. Figure~6 shows the Fe/O and Ne/O ratios as 
a function of the average temperature during quiescence, flare rise, and flare peak.

\begin{figure}[!t]
\includegraphics[width=5.5cm]{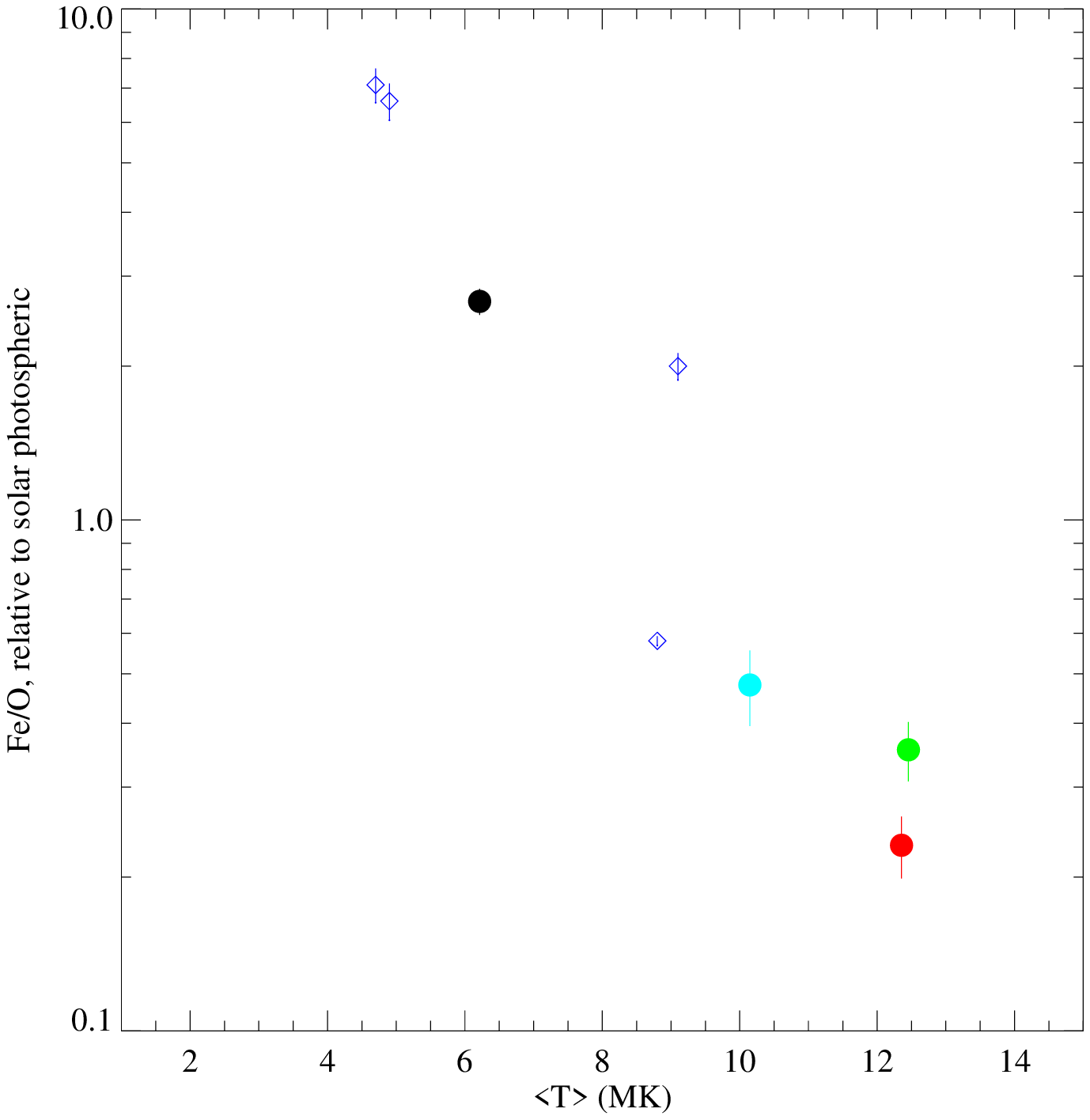}
\includegraphics[width=5.5cm]{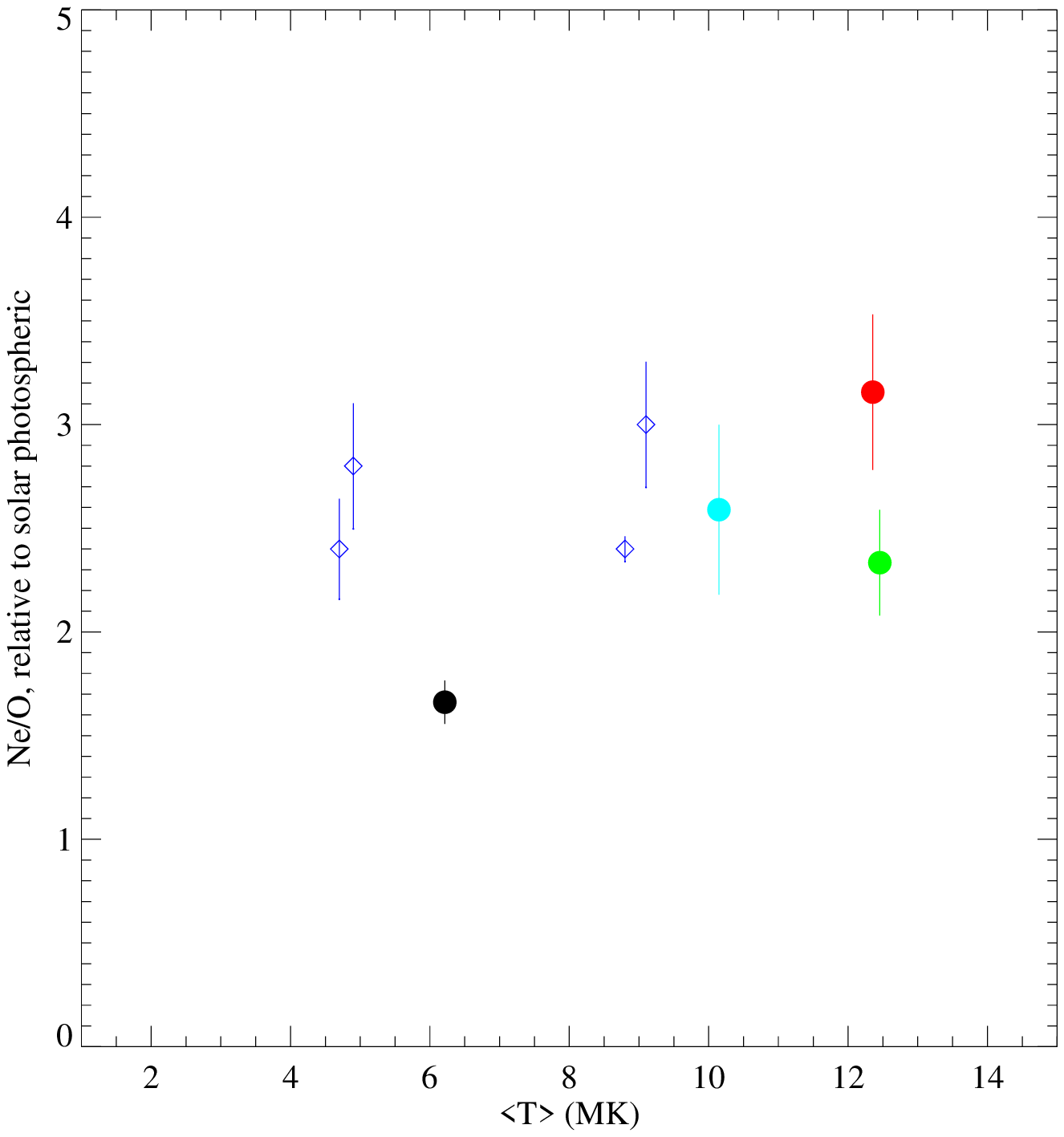}
\caption{Coronal abundance ratio as a function of $<T>$. 
\textit{Left}: Fe/O ratios relative to solar photospheric for RS CVn 
binary systems (dots) with similar ratios for solar analogs (open 
diamonds, from G\"udel et al. 2002; these proceedings). Note the logarithmic 
vertical scale. \textit{Right}: Similar but for Ne/O ratios.}
\end{figure}

\begin{figure}[!t]
\includegraphics[width=5.5cm]{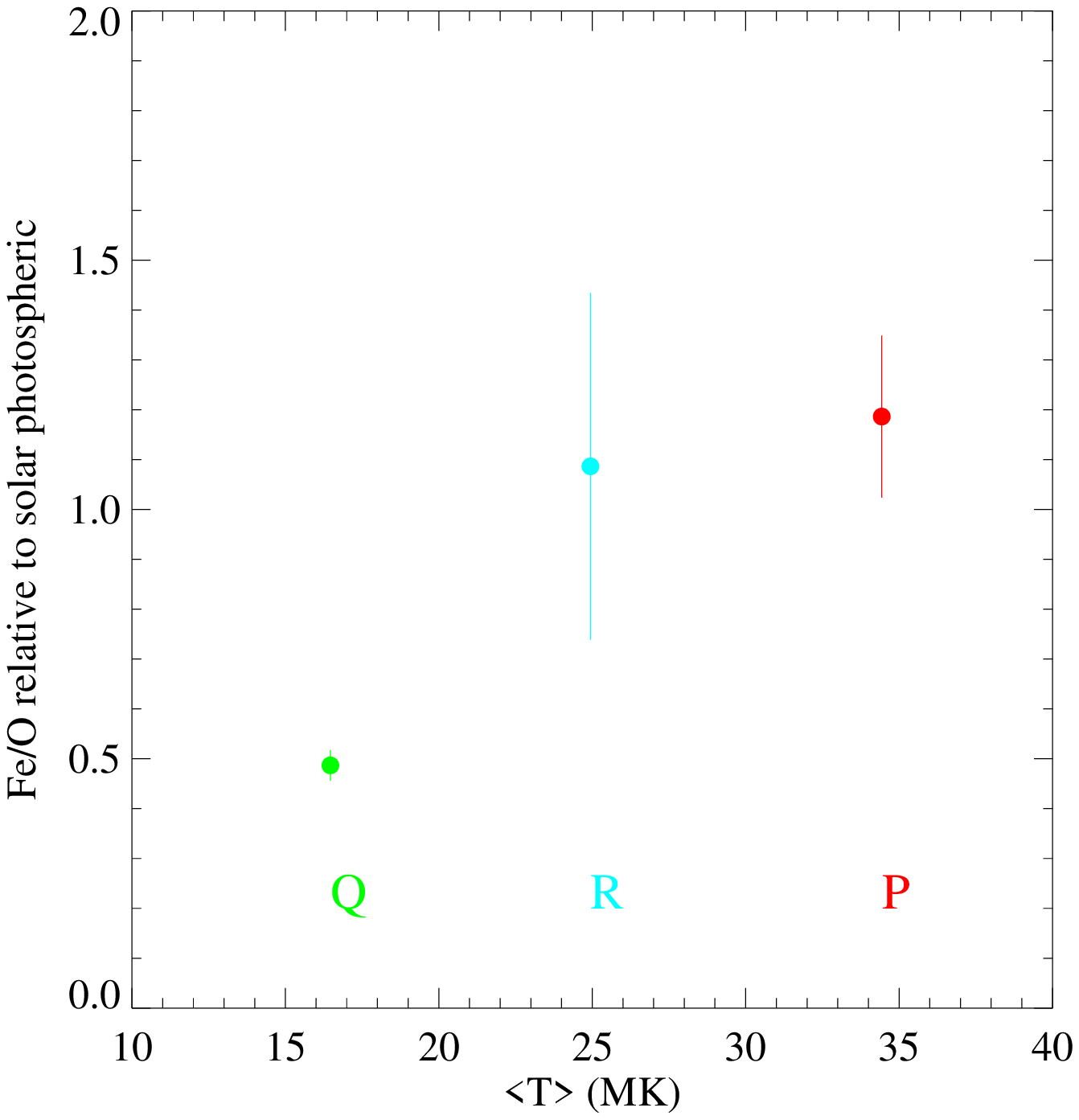}
\includegraphics[width=5.5cm]{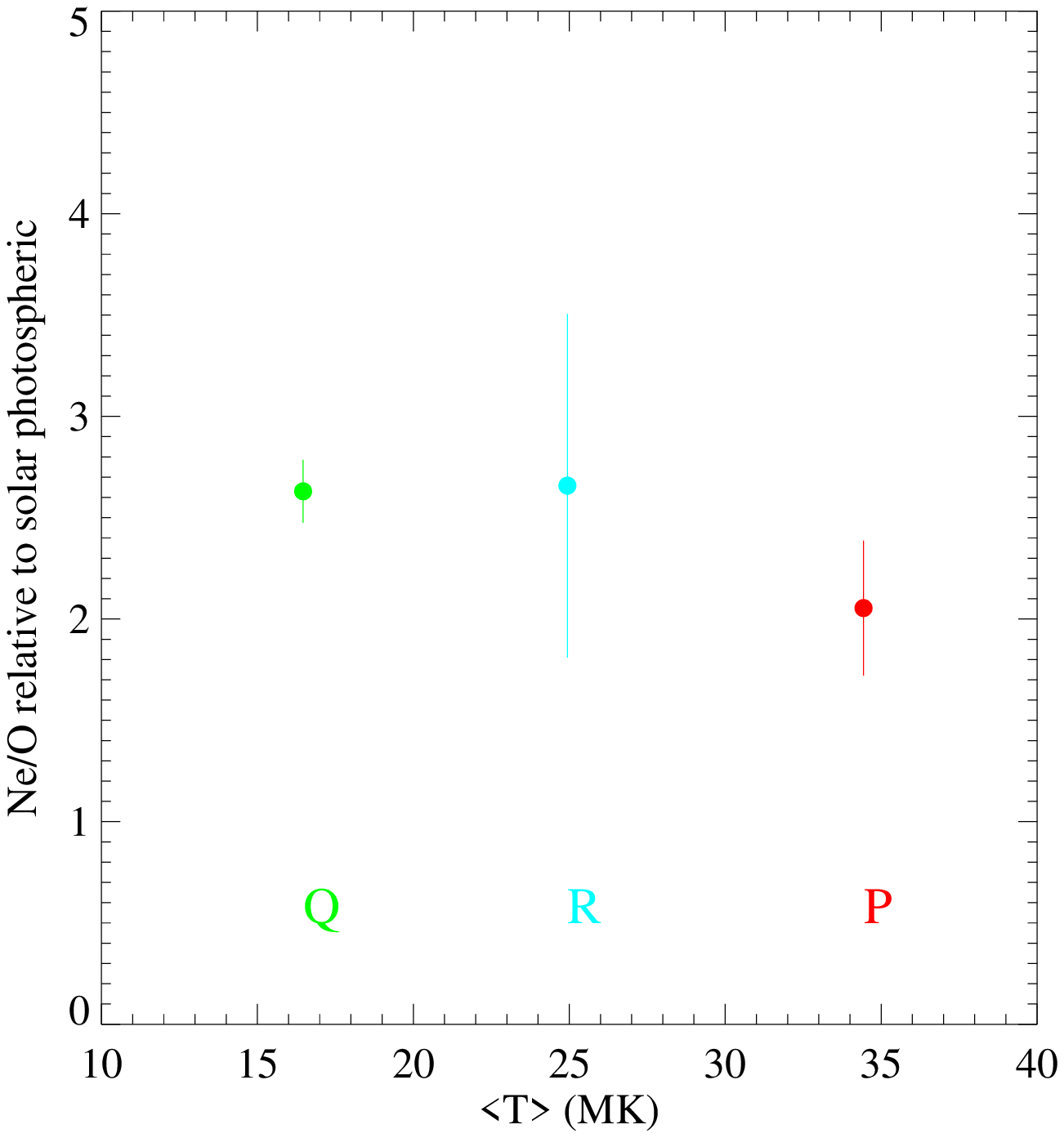}
\caption{Coronal abundance ratio as a function of $<T>$ during
a large flare in HR~1099. Left panel give Fe/O ratios, while the right panel 
gives Ne/O ratios. `Q' stands for quiescent, `R' for flare rise, and `P' for 
flare peak.}
\end{figure}

Consistent with the previous analysis (Audard et al. 2001b) and with previous 
results from other observations, the absolute Fe abundance increases during the rising phase of the flare; 
however, the absolute Ne abundance stays constant. The Fe/O ratio also 
increases, while Ne/O remains constant, consistent with low-FIP elements being 
enhanced during flares while high-FIP elements staying equally abundant. 
Other low-FIP elements and high-FIP elements show similar respective behaviors.

\section{Discussion and Conclusion}

We have found a correlation between the coronal elemental composition
in active stars and the level of magnetic activity. The very active RS
CVn binary systems show a clear trend toward an enhancement of coronal 
abundances of high-FIP elements (e.g., C, N, O, Ne) relative to low-FIP 
elements (e.g., Fe, Mg, Si). It is interpreted as a further evidence for an
``Inverse First Ionization Potential'' effect as found by Brinkman et al.
(2001) in a deep exposure of HR~1099. Our sample (HR~1099, UX
Ari, $\lambda$ And, and Capella) covers high to intermediate magnetic activity 
levels.  The most active stars show a clear IFIP effect, while the intermediately 
Capella show either an absence of a FIP bias or a weak FIP effect
(Fig.~3). Using the average coronal temperature as an activity
indicator, we find that coronal abundances of low-FIP elements (normalized to O)
decrease with increasing temperature, while they stay constant for high-FIP 
elements. Note that such behavior will change if abundances are 
normalized to a low-FIP element such Fe. However, abundances relative to H
suggest that low-FIP elemental abundances do vary with the level of magnetic
activity, while this is not the case for high-FIP elements. 

Our results agree well with the long-term evolution from an inverse FIP
effect to a ``normal'' FIP effect found in solar analogs of solar photospheric
composition (G\"udel et al. 2002; these proceedings). 
Unfortunately, photospheric abundances are unknown for most RS CVn binary systems,
and if known, measurements are largely scattered because it is difficult
to obtain reliable photospheric abundances from optical spectroscopy, mainly
because of enhanced chromospheric activity, high rotation velocities, and the
presence of spots. Our target $\lambda$ And is one of the rare cases with
several measured photospheric abundances. Coronal abundances relative to the
photospheric abundances derived by Donati et al. (1995) show no clear correlation
with the FIP (Fig.~4). Accurate measurements of 
photospheric abundances in the most active stars are timely.

\acknowledgments

M.~A. acknowledges support from the Swiss National 
Science Foundation (grant 2100-049343). He also thanks the
organizers of the CS12 conference for financial
support. The Space Research Organization of the Netherlands (SRON) is supported 
financially by NWO. This work is based on observations obtained with XMM-Newton, an ESA science 
 mission with instruments and contributions directly funded by ESA Member 
 States and the USA (NASA).

\index{*V711 Tau| see {HR 1099}}
\index{*$\alpha$ Aur|see {Capella}}
\index{inverse FIP|see {abundances}}
\index{FIP|see {abundances}}

\end{document}